\documentclass[aps,pra,notitlepage,nofootinbib]{revtex4-1}
\usepackage[latin9]{inputenc}
\usepackage{amsmath}
\usepackage{graphicx}

\usepackage{bbm}
\usepackage{amsfonts}
\usepackage{times}

\def\be{\begin{equation}}\def\ee{\end{equation}}\def\lb{\label}

\begin{document}

\title{Physical velocity of particles in relativistic curved momentum space}

\author{Salvatore Mignemi}

\affiliation{Dipartimento di Matematica e Informatica, Universit\`a di Cagliari\\
viale Merello 92, 09123 Cagliari, Italy\\
 and INFN, Sezione di Cagliari, Cittadella Universitaria, 09042 Monserrato, Italy}

\author{Giacomo Rosati}

\affiliation{Institute
for Theoretical Physics, University of Wroc\l{}aw, Pl.\ Maksa Borna
9, Pl--50-204 Wroc\l{}aw, Poland}
\begin{abstract}
We show in general that for a relativistic theory with curved momentum
space, i.e.~a theory with deformed relativistic symmetries, the physical
velocity of particles coincides with their group velocity. This clarifies
a long-standing question about the discrepancy between
coordinate and group velocity for this kind of theories. The first
evidence that this was the case had been obtained at linear order in the deformation
parameter in \cite{RLspeedJackNicco} for the specific case of $\kappa$-momentum space.
The proof was based on the recent
understanding of how relative locality affects these scenarios. We
here rely again on a careful implementation of relative locality effects,
and obtain our result for a generic (relativistic) curved momentum
space framework at all orders in the deformation/curvature parameter.
We also discuss the validity of this result when the deformation depends on the
coordinates as well as on the momenta.
\end{abstract}
\maketitle

\section{Introduction}

The possibility that momentum space of relativistic particles may
be curved~\cite{RelLocPrinciple,JurekReview,ArzanoQfieldCurved,CarmonaRelLoc,anatomy,FlaGiukRelLoc,FreidelFieldCurved,DualRedshift,Palmisano,MigSamSnyderRL,GiuliaBoosts} has gained a significant attention especially
in studies of DSR (``doubly'', or for some authors ``deformed'',
special relativity) theories~\cite{GACDSR,JurekDSR,SmolMagDSR}. In these
theories relativistic symmetries are deformed at an observer-independent
momentum scale $\kappa$, which in the quantum gravity literature is typically taken to be close to the Planck scale,
$\kappa\sim E_{P}/c=\sqrt{\hbar c^{3}/G}\sim{10}^{-19}\text{GeV}/c$.
These theories are particularly relevant because they offer the possibility, at least in some cases, of testing
Planck scale effects, related to quantum gravity phenomenology~\cite{GACphen,GACGRB,DSRFRW,IceCubeNature},
modifying the propagation of relativistic particles.

It has been recently understood~\cite{RLspeedJackNicco,RelLocPRL,RelLocPrinciple} that an observer-independent
momentum-scale deformation of relativistic symmetries
implies the notion of absolute locality to be relaxed in favour of
a ``relative locality'', a sort of ``momentum space redshift'',
denoted as ``lateshift'' in~\cite{DualRedshift}.
In fact, one can understand these features to result from the momentum space of particles to be ``curved'',
with $\kappa$ the scale of curvature~\cite{JurekReview,anatomy,FlaGiukRelLoc,multipart}. The idea of curved momentum space dates back
to the seminal work of Born~\cite{BornReciprocity}, and Snyder \cite{Snyder} and has been later
more rigorously formalized in the context of Hopf-algebras~\cite{MajidFoundation}
as dual to non-commutativity of spacetimes coordinates. It is worth
to point out that in order to still have a full set of (deformed)
relativistic symmetries, the possible choices of curved momentum spaces
is substantially restricted, essentially to maximally symmetric spaces.

For what regards the study of (free) particle motion in this context,
the main features of momentum space curvature and spacetime non-commutativity
can be captured in the characterization of the phase space of particles.
In the ``relative-locality'' regime~\cite{RelLocPrinciple}, $\hbar\rightarrow0$
and $G\rightarrow0$ but $\hbar/G\neq0$, where one neglects $\hbar$-driven
quantum corrections, the particle phase space can be described in
terms of Poisson brackets $\left\{ \cdot,\cdot\right\} $ between
momenta $p_{\mu}$ and spacetime coordinates $x^{\mu}$, the residual
deformation being encoded in the constant scale $\kappa$.

The notion of particle velocity in the DSR context has been a matter
of debate for some time~\cite{lukieANNALS,gacmaj,jurekvelISOne,kowaVEL1,japaVEL,lukieVEL,kosinskiVEL,gacMandaniciDANDREA,mignemiVEL,grilloVEL,ghoshVELisONE,ghoshVEL}, and has found an assessment
mainly due to the understanding of relative locality~\cite{RLspeedJackNicco}.
The question is that there is in general a difference between the
coordinate velocity $v_{\text{kin}}^{j}=dx^{j}/dt$ obtained from
Hamiltonian analysis and the group velocity $v_{kin}^{j}=\partial p_{0}(p_{j})/\partial p_{j}$
coming directly from the dispersion relation. It was shown in~\cite{RLspeedJackNicco}
for the case of $\kappa$-momentum space, and then successfully applied
to several following studies~\cite{anatomy,causality,DSRFRW,DualRedshift} based on the same
framework, that if one takes into account relative locality by considering
the effect of deformed translations, the two notions of velocity give rise to the
same physical predictions, the group velocity representing the actual ``physical'' velocity.

However, the resolution of the velocity issue has been studied so
far only for some particular cases of interest, and only at first
order in the deformation/curvature parameter (except for the Snyder model, where the analysis was performed at all orders in \cite{MigSamSnyderRL}). 
In this letter we address the goal of characterizing the physical velocity of particles in a general
scenario of deformed relativistic symmetries or associated (relativistic)
curved momentum space. We obtain that, for a deformation depending only on momenta, the identification of the
physical velocity with the group velocity is true in general, at all orders in
the deformation parameter.
Some care must be taken when the deformation depends also on the spacetime coordinates.
In this case, no general answer can be obtained, except when some conditions are imposed on the
form of the Poisson brackets of the theory.

\section{Physical velocity in (relativistic) curved momentum space}

\subsection{Construction of phase space from momentum space metric\label{sec:CurvedPhaseSpace}}

For our arguments, it is sufficient to consider the Hamilton equations obtained from the deformed
symplectic structure and the Hamiltonian, see next section. However, we summarize here
for completeness the construction of the action for a
relativistic particle with curved momentum space associated to a non-commutative
spacetime. We consider only the case in which the curved momentum
space is maximally symmetric. It is easy to show that in this case
the momentum space curvature can be considered to be dual to some
non-commutative spacetime coordinates (typically of Lie algebra type).

Indeed, in a theory with curved momentum space with a metric
\begin{equation}
dp^{2}=h^{\mu\nu}\left(p\right)dp_{\mu}dp_{\nu},\label{MomSpaceMetric}
\end{equation}
where
\begin{equation}
h^{\mu\nu}\left(p\right)=\eta^{\mu\nu}+O\left(\frac{p}{\kappa}\right),
\end{equation}
maximal symmetry implies the existence of four Killing vectors $\xi_{\nu}^{\mu}\left(p\right)=\delta_{\nu}^{\mu}+O\left(p/\kappa\right)$
associated to ``translations in momentum space''. Considering vectors
$\chi^{\mu}$ tangent to momenta $p_{\mu}$, one can define canonical
Poisson brackets
\begin{equation}
\left\{ \chi^{\mu},\chi^{\nu}\right\} =0,\qquad\left\{ p_{\mu},\chi^{\nu}\right\} =\delta_{\mu}^{\nu},\qquad\left\{ p_{\mu},p_{\nu}\right\}=0.\label{canonicalPoisson}
\end{equation}
Then, a natural definition of deformed phase space arises by taking non-commutative coordinates $x^\mu$ such that
\begin{equation}
x^{\mu}=\xi_{\nu}^{\mu}\left(p\right)\chi^{\nu},\quad \chi^{\mu}=\bar{\xi}_{\nu}^{\mu}\left(p\right)x^{\nu},\label{nonCommCoord}
\end{equation}
where $\xi_{\nu}^{\mu}\left(p\right)$ is assumed to be invertible,
$\bar{\xi}_{\nu}^{\mu}\left(p\right)$ being its inverse. From (\ref{canonicalPoisson})
we get
\begin{equation}
\left\{ x^{\mu},x^{\nu}\right\} =\left(\xi_{\rho}^{\nu}\left(p\right)\frac{\partial\xi_{\sigma}^{\mu}\left(p\right)}{\partial p_{\rho}}-\xi_{\rho}^{\mu}\left(p\right)\frac{\partial\xi_{\sigma}^{\nu}\left(p\right)}{\partial p_{\rho}}\right)\bar{\xi}_{\tau}^{\sigma}\left(p\right)x^{\tau}=\psi_{\rho}^{\mu\nu}\left(p\right)x^{\rho},
\label{noncommutative}
\end{equation}
\begin{equation}
\left\{ p_{\mu},x^{\nu}\right\} =\xi_{\mu}^{\nu}\left(p\right).
\end{equation}
Non-commutative coordinates (\ref{nonCommCoord}) are the generators
of ``translations in momentum space'', fulfilling the notion of
non-commutativity/curvature duality from the Hopf-algebraic point
of view.

From this choice of coordinates, the kinetic term for the particle
Lagrangian arises naturally as
\begin{equation}
{\cal L}_{kin}=\chi^{\mu}\dot{p}_{\mu}=x^{\nu}\bar{\xi}_{\nu}^{\mu}\left(p\right)\dot{p}_{\mu}.
\end{equation}
The free particle action is then
\begin{equation}
S=\int ds\ \left(x^{\nu}\bar{\xi}_{\nu}^{\mu}\left(p\right)\dot{p}_{\mu}+\lambda{\cal {\cal H}}\left(p\right)\right),\label{action}
\end{equation}
where $\lambda$ is a Lagrange multiplier and the Hamiltonian for
a free particle of mass $m$ is
\begin{equation}\lb{ham}
{\cal H}={\cal C}\left(p\right)-m^{2}.
\end{equation}
Here ${\cal C}\left(p\right)$ is a function of the momenta corresponding
to the mass Casimir of the algebra of symmetries associated to the
metric (\ref{MomSpaceMetric}) (see for instance~\cite{anatomy,multipart}), such that
\begin{equation}
{\cal C}\left(p\right)=\eta^{\mu\nu}p_{\mu}p_{\nu}+O\left(\frac{1}{\kappa}\right),
\end{equation}
and ${\cal C}_{0}=\eta^{\mu\nu}p_{\mu}p_{\nu}$ is the standard special
relativistic mass Casimir. 

The correspondance of the Hamiltonian with the (deformed) mass Casimir ensures that the action is invariant under the action of the deformed relativistic transformations generated, by Poisson brackets, by the charges associated to space and time translations, rotations, and boosts.
Particularly important for our argument are translational symmetries, and we identify as usual the momenta $p_\mu = (p_0,p_j)$ as the generators of time and space translations.

${\cal C}\left(p\right)$ can
be also calculated from the invariant function corresponding to the
geodesic length (see~\cite{anatomy}) in momentum space
\begin{equation}
{\cal C}\left(p\right)=\int_{0}^{1}ds\ h^{\mu\nu}\left(p\right)\dot{\gamma}_{\mu}\dot{\gamma}_{\nu},\qquad\gamma_{\mu}\left(0\right)=0,\quad\gamma_{\mu}\left(1\right)=p_{\mu}.
\end{equation}

The variation of the action (\ref{action}) gives
\begin{equation}\lb{variation}
\delta S=\int ds\ \left(\delta x^{\nu}\bar{\xi}_{\nu}^{\mu}\left(p\right)\dot{p}_{\mu}+\delta p_{\mu}\left(\lambda\frac{\partial{\cal C}\left(p\right)}{\partial p_{\mu}}-\bar{\xi}_{\nu}^{\mu}\left(p\right)\dot{x}^{\nu}+x^{\nu}\dot{p}_{\rho}\left(\frac{\partial\bar{\xi}_{\nu}^{\rho}\left(p\right)}{\partial p_{\mu}}-\frac{\partial\bar{\xi}_{\nu}^{\mu}\left(p\right)}{\partial p_{\rho}}\right)\right)+\delta\lambda{\cal H}\right)
\end{equation}

\subsection{Coordinate velocity}

Starting from the previous considerations, we study a phase space with Poisson brackets 
\begin{equation}\lb{sympl}
\left\{ x^{\mu},x^{\nu}\right\} =\psi_{\rho}^{\mu\nu}\left(p\right)x^{\rho},\qquad\left\{ p_{\mu},x^{\nu}\right\} =\xi_{\mu}^{\nu}\left(p\right),\qquad\left\{ p_{\mu},p_{\nu}\right\} =0.
\end{equation}
This  general form comprises most noncommutative geometries considered in DSR literature: $\kappa$-Minkowski \cite{lukieANNALS}, Magueijo-Smolin \cite{SmolMagDSR}, Snyder \cite{Snyder}, etc.

The Hamilton equations following from the symplectic structure \eqref{sympl} and from the Hamiltonian \eqref{ham}, or equivalently from the variation \eqref{variation} are
\begin{equation}
\dot{x}^{\mu}=\lambda\,\xi_{\nu}^{\mu}\left(p\right)\frac{\partial{\cal C}\left(p\right)}{\partial p_{\nu}},
\label{EqMotionx}
\end{equation}
\begin{equation}
\dot{p}_{\mu}=0.
\label{EqMotionp}
\end{equation}
The coordinate (or kinematic) velocity can be defined in a natural way as
\begin{equation}
v_{\text{kin}}^{j}\left(p\right)=\frac{dx^{j}\left(x^{0}\right)}{dx^{0}}=\frac{\dot{x}^{j}}{\dot{x}^{0}}\Big|_{{\cal C}\left(p\right)=m^{2}}=\frac{\xi_{\mu}^{j}\left(p\right)\partial{\cal C}\left(p\right)/\partial p_{\mu}}{\xi_{\nu}^{0}\left(p\right)\partial{\cal C}\left(p\right)/\partial p_{\nu}}\Big|_{{\cal C}\left(p\right)=m^{2}}.\label{VelKin}
\end{equation}
Because of \eqref{EqMotionp}, the velocity is constant and the trajectories of the particles are straight lines.

However, in order to consistently define the notion of physical velocity, we have
to take into account relative locality and consider the effect
of deformed (spacetime) translations, generated by the action of the momenta $p_{\mu}$ on the coordinates $x^{\mu}$
\cite{RLspeedJackNicco}. Consider an infinitesimal translation. This
is defined by
\begin{equation}
{x'}^{\mu}=x^{\mu}-\epsilon^{\nu}\left\{ p_{\nu},x^{\mu}\right\} =x^{\mu}-\epsilon^{\nu}\xi_{\nu}^{\mu}\left(p\right).\label{TranInf}
\end{equation}
One can check, as expected from the discussion above, that the action, and thus the symplectic structure~\footnote{In particular one can check that the ``non-commutative'' spacetime coordinate Poisson brackets~(\ref{noncommutative}) are preserved under deformed translations: $\left\{ {x'}^{\mu},{x'}^{\nu}\right\} =\psi_{\rho}^{\mu\nu}\left(p\right){x'}^{\rho}$.}, is left invariant under deformed translations of this kind.

Let us first discuss the usual argument for the definition of the physical velocity used in relative locality
literature~\cite{RLspeedJackNicco,anatomy}, valid at first order in the deformation
parameter $1/\kappa$.

\subsection{The usual argument in relative locality revisited: worldlines and
deformed finite translations.}

Since momenta (Poisson-) commute, it is trivial
to obtain a finite translation, by iteration of infinitesimal translations
(\ref{TranInf}), resulting simply in
\begin{equation}
{x'}^{\mu}=x^{\mu}-a^{\nu}\left\{ p_{\nu},x^{\mu}\right\} =x^{\mu}-a^{\nu}\xi_{\nu}^{\mu}\left(p\right).\label{finiteTrans}
\end{equation}
A particle (coordinate) trajectory can be calculated to be\footnote{Notice that since the momenta are conserved (cf.~\eqref{EqMotionp}),
the velocity \eqref{VelKin} does not depend on time.}
\begin{equation}
x^{j}\left(x^{0}\right)=\bar{x}^{j}+v_{\text{kin}}^{j}\left(p\right)\left(x^{0}-\bar{x}^{0}\right).\label{trajectory}
\end{equation}
To calculate the physical velocity, we consider two distant
observers at relative rest, local respectively to the emission (observer
$A$) and detection (observer $B$) of the particle. This means that
the trajectory described by the translated observer $B$ will again be given
by (\ref{trajectory}), with primed coordinates, and initial
conditions such that the particle has been emitted at the first observer's
$A$ origin. From (\ref{finiteTrans}) we find
\begin{equation}
\bar{x}'^{\mu}=-a^{\nu}\xi_{\nu}^{\mu}\left(p\right).
\end{equation}
Now, we impose the condition for which the
translated observer $B$ is sitting along the worldline of a soft
particle (for which the effects of deformation can be neglected),
emitted at the origin of $A$ simultaneously with a hard particle.
This amounts to set
\begin{equation}
a_{s}^{j}=a_{s}^{0}v_{0}^{j},
\end{equation}
where\footnote{Notice that in our notation, using covariant indexes
with Lorentzian metric, $p_{j}<0$ for $v^{j}>0$; we take positive
energy solutions $p_{0}=\left|\vec{p}\right|+O\left(\frac{1}{\kappa}\right)>0$.}
\begin{equation}
v_{0}^{j}=-\frac{p_{j}}{p_{0}}\Big|_{{\cal C}_{0}=m^{2}}=-\frac{p_{j}}{\sqrt{\left|p\right|^{2}+m^{2}}}.
\end{equation}
We find the worldline of the hard particle described by the translated observer
\begin{equation}
{x'}^{j}\big({x'}^{0}\big)=-a_{s}^{\mu}\xi_{\mu}^{j}\left(p\right)+v_{\text{kin}}^{j}\left(p\right)\left({x'}^{0}+a_{s}^{\mu}\xi_{\mu}^{0}\left(p\right)\right).
\end{equation}
Setting to zero ${x'}^{j}\big({x'}^{0}\big)$, we obtain the time
of detection as the time at which the hard particle crosses the spatial
origin of the translated observer:
\begin{equation}\lb{xdet}
{x'}_{\text{\ensuremath{\det}}}^{0}=\frac{a_{s}^{\mu}\xi_{\mu}^{j}\left(p\right)-v_{\text{kin}}^{j}\left(p\right)a_{s}^{\mu}\xi_{\mu}^{0}\left(p\right)}{v_{\text{kin}}^{j}\left(p\right)}
\end{equation}

\begin{figure}[h!]
\centering{}\includegraphics[scale=0.7]{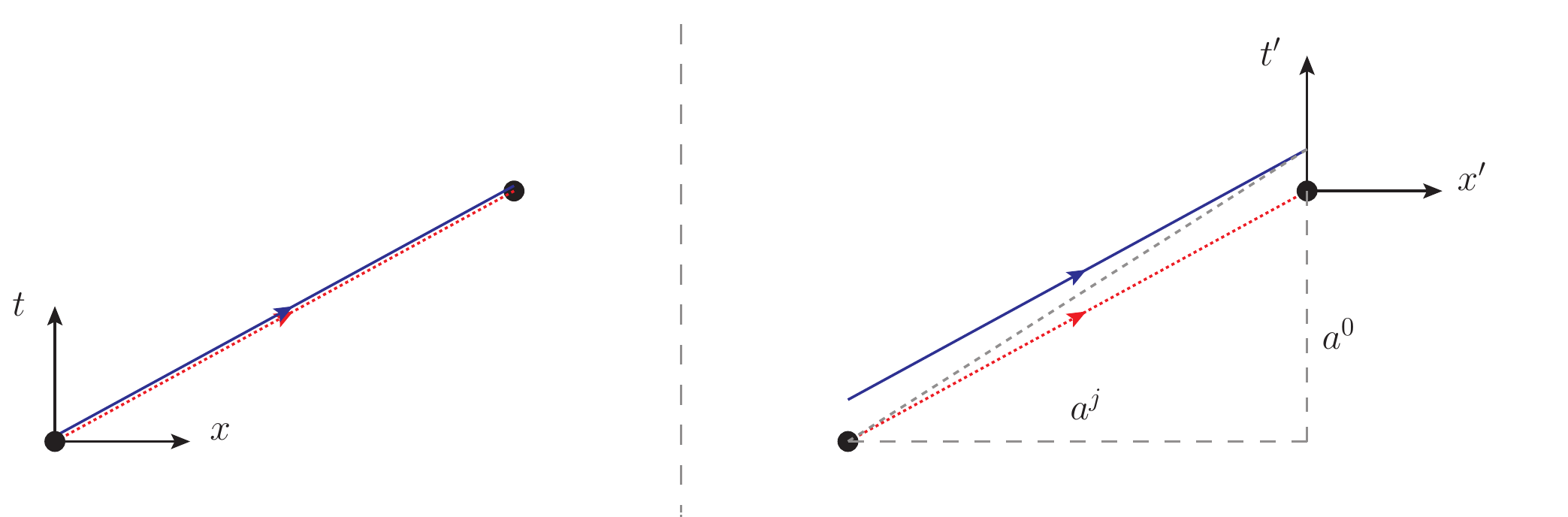} \caption{\label{fig:RelLocVel}We here show a graphical depiction of the usual
relative locality argument for the case of bicrossproduct $\kappa$-momentum
space (see~\cite{RLspeedJackNicco,anatomy}). The left and right panels are respectively
the observers $A$ and $B$ description of the particles worldlines
($t=x^{0}$). A hard particle (continuous blue worldline) and a soft
particle (dotted red worldline), are emitted simultaneously at $A$'s
spatial origin. The observer $B$ is sitting at the detection event
of the soft particle, while the hard particle is described by
$B$ to reach its spatial origin its spatial origin at the time ${x'}_{\text{\ensuremath{\det}}}^{0}\simeq\frac{1}{\kappa}a_{s}^{0}\left|\vec{p}\right|$.
In dashed grey is what is assumed to be the physical velocity.}
\end{figure}

We can solve \eqref{xdet} at first order in $1/\kappa$ to find that in
general the hard particle does not arrive at the observer $B$'s origin
(see Fig.~\ref{fig:RelLocVel}). For instance it has been shown~\cite{RLspeedJackNicco,anatomy}
that in the case of $\kappa$-momentum space corresponding to ``time-to-the-right''
bicrossproduct basis of $\kappa$-Poincar\'e, where
\begin{equation}
\begin{gathered}{\cal C}\simeq p_{0}^{2}-\vec{p}^{2}+\frac{1}{\kappa}p_{0}\vec{p}^{2},\\
\xi_{\mu}^{\nu}\simeq\delta_{\mu}^{\nu}+\frac{1}{\kappa}\delta_{0}^{\nu}\delta_{\mu}^{j}p_{j},
\end{gathered}
\end{equation}
for massless particles
$p_{j}<0$ for $v^{j}>0$)
\begin{equation}
v_{\text{kin}}^{j}=-\frac{p_{j}}{\left|\vec{p}\right|}=v_{0}^{j},
\end{equation}
and the shift in time detection between the arrival of the hard and soft particles
(the time delay) is
\begin{equation}
{x'}_{\text{\ensuremath{\det}}}^{0}\simeq\frac{1}{\kappa}a_{s}^{0}\left|\vec{p}\right|.
\end{equation}
As it was first noticed in~\cite{RLspeedJackNicco}, the time shift
is the one that would have been obtained if the ``physical'' velocity were
\begin{equation}
v_{\text{phys}}^{j}\left(p\right)=-\frac{p_{j}}{\left|\vec{p}\right|}\left(1-\frac{1}{\kappa}\left|\vec{p}\right|\right)=\frac{dp_{0}\left(p\right)}{dp_{j}}=\frac{\partial{\cal C}/\partial p_{j}}{\partial{\cal C}/\partial p_{0}}\Big|_{{\cal C}\left(p\right)=0}.
\end{equation}
Thus, at least at first order in $1/\kappa$, and at least for the
case of $\kappa$-momentum space here considered, even if the coordinate
velocity of a massless particle is energy-independent, once the relative locality effects arising from the deformed translations have been
taken into account, the physical velocity coincides with the group velocity
$v_{\text{phys}}^{j}\left(p\right)$, and for nontrivial ${\cal C}$ a time-delay effect occurs.

\subsection{Physical velocity is the group velocity: a simple general argument}

We can give a general characterization of the physical velocity, to
all orders in the deformation parameter, starting from the following
considerations. The notion of physical
velocity stemming from the implementation of relative locality, i.e.~the
one implied by the derivation of the previous section, can be
understood as the slope (in spacetime) of the physical worldline,
consisting in the worldline composed by the set of observers, at relative
rest, that actually sit on the particle worldline. To be more precise,
one can think of fixing a common spacetime background for observers
at relative rest by mutual exchange of low energy (soft) photons. Each
point in this spacetime can be thought as the (spacetime) origin
of one of these observers. Then, the physical worldline of the particle 
is represented by the points in this spacetime joining the origins of
all the observers actually crossed by the particle.

The situation is more clear if one considers the example depicted in
Fig.\ref{fig:RelLocVel}. Instead of considering a second observer
translated along the soft photon worldline, we can perform a set of
infinitesimal translations, such that each observer still sits on
the hard worldline. After each infinitesimal translation the hard
particle coordinate worldline is shifted rigidly by a certain amount
along the physical worldline that we want to find. It is important
to notice that this is so because the covariance of the formalism,
ensured by the Jacobi relations, is such that all observers describe
the same coordinate velocity $v_{\text{kin}}^{j}\left(p\right)$,
i.e. the same slope for particle worldlines having the same mass,
and that translations do commute, so that a translation does not change
the slope of the worldline, i.e.~the coordinate velocity. What we
will reconstruct joining the origins of the observers obtained in
this way, will be the dashed gray line in Fig.\ref{fig:RelLocVel}.
This is the physical worldline, the (spacetime) slope of which will
be the physical velocity.

\begin{figure}[h!]
\centering{}\includegraphics[scale=0.7]{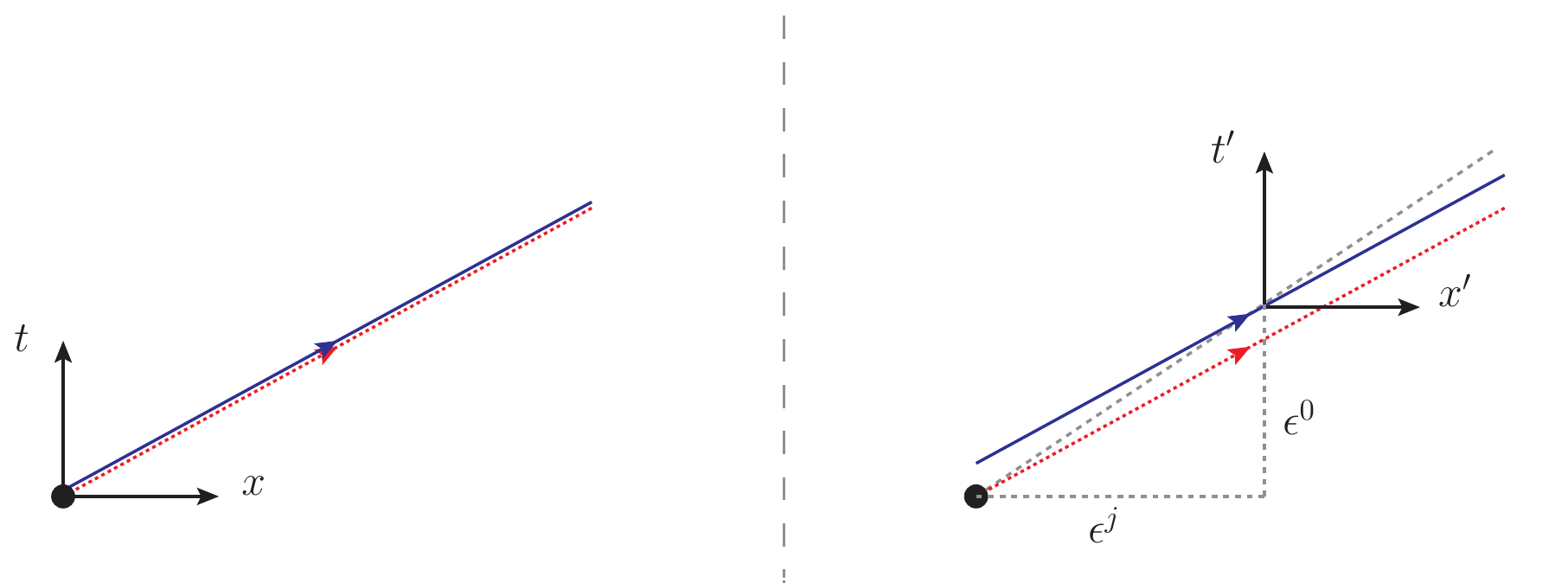} \caption{\label{fig:RelLocPhysVel} Here in right panel the (infinitesimally)
translated observer is sitting on the hard particle worldline. The
coordinate worldline is shifed rigidly along the physical worldline
(dashed gray), so that they still coincide in the translated observer's
origin.}
\end{figure}

Thus, in order to obtain the physical velocity, it is enough to consider
infinitesimal translations. What we have to do is to ask ourselves
what are the coordinates of an observer translated infinitesimally
along the hard worldline, such that it still detects the hard worldline
in its origin. The situation is depicted in Fig. \ref{fig:RelLocPhysVel}.

Now, the conditions to impose on the translated coordinates are two: the
first is that they belong to the (translation of the) coordinate worldline
described by the first observer $A$. This is given by setting the relation
\begin{equation}
x^{j}=v_{\text{kin}}^{j}\left(p\right)x^{0},
\end{equation}
in (\ref{TranInf}). The second is that the point of the worldline
corresponding to its crossing with the translated observer's origin
has translated coordinates equal to zero, so that for this point,
we get from (\ref{TranInf}) a condition on the translation parameters:
\begin{equation}
\epsilon^{\mu}=\bar{\xi}_{\nu}^{\mu}\left(p\right)x^{\nu}.
\end{equation}
The physical velocity will then be given by the slope (see Fig. \ref{fig:RelLocPhysVel})
$\epsilon^{j}/\epsilon^{0}$, which, using these two conditions, becomes
\begin{equation}
\begin{split}v_{\text{phys}}^{j}\left(p\right)= & \frac{\epsilon^{j}}{\epsilon^{0}}=\frac{\bar{\xi}_{\mu}^{j}\left(p\right)x^{\mu}}{\bar{\xi}_{\mu}^{0}\left(p\right)x^{\mu}}\\
= & \frac{\bar{\xi}_{0}^{j}+\bar{\xi}_{i}^{j}v_{\text{kin}}^{i}}{\bar{\xi}_{0}^{0}+\bar{\xi}_{i}^{0}v_{\text{kin}}^{i}}\\
= & \frac{\bar{\xi}_{0}^{j}\dot{x}^{0}+\bar{\xi}_{i}^{j}\dot{x}^{i}}{\bar{\xi}_{0}^{0}\dot{x}^{0}+\bar{\xi}_{i}^{0}\dot{x}^{i}}\Bigg|_{{\cal C}=m^{2}},
\end{split}
\end{equation}
where in the last step we used Eq. (\ref{VelKin}). But now, using
the equations of motion (\ref{EqMotionx}), the last relation is just
\begin{equation}
v_{\text{phys}}^{j}\left(p\right)=\frac{\left(\bar{\xi}_{0}^{j}\xi_{\mu}^{0}+\bar{\xi}_{i}^{j}\xi_{\mu}^{i}\right)\partial{\cal C}\left(p\right)/\partial p_{\mu}}{\left(\bar{\xi}_{0}^{0}\xi_{\nu}^{0}+\bar{\xi}_{i}^{0}\xi_{\nu}^{i}\right)\partial{\cal C}\left(p\right)/\partial p_{\nu}}\Bigg|_{{\cal C}=m^{2}}=\frac{\partial{\cal C}/\partial p_{j}}{\partial{\cal C}/\partial p_{0}}\Big|_{{\cal C}=m^{2}}=\frac{dp_{0}\left(p\right)}{dp_{j}}.
\end{equation}
Notice that it is crucial for this derivation that the phase space
terms $\xi_{\mu}^{\nu}\left(p\right)$ are invertible. We obtain in
this way the sought result. The physical velocity coincides with the
group velocity obtained from the deformed mass Casimir.

\subsection{Generalization to coordinate-dependent Poisson brackets}
More general models can be envisaged in which the Poisson brackets $\{p_\mu,x^\nu\}$ depend on the spacetime coordinates, as well as on
the momenta.  This occurs for example in a recently proposed model based on the quantum clock \cite{MignemiUras}.
In this case our proof is not valid and the two definitions of velocity might give rise to different predictions.
In fact, the argument of the previous section relies on the fact that under the given hypotheses the trajectories are straight lines and the
velocity is constant. Some conditions must therefore be satisfied in order for this to be true.
Here we assume, as before, that $\{p_\mu,p_\nu\}=0$ and that the Hamiltonian depends only on the momenta $H=H(p)$.
This excludes the case of curved spacetimes, which should be investigated separately.

The conditions on the trajectories can be satisfied if the Poisson brackets of the coordinates and momenta have a Lorentz covariant form:
\be\lb{conditions}
\{p_\mu,x^\nu\}=\xi^\nu_\mu=f(p^2,x\cdot p\,,x^2)\delta_{\mu\nu}+g(p^2,x\cdot p\,,x^2)p_\mu p_\nu,
\ee
while $\{x_\mu,x_\nu\}$ is some function of $x$ and $p$ determined by the Jacobi identities.
This is a slight extension of the generalized Snyder models introduced in \cite{Meljanac,MelMig}.
In fact, in this case, using the Hamilton equations \eqref{EqMotionx}, the ratios
\be
\frac{\dot{x}^\mu}{\dot{x}^\nu}=\frac{\xi^\mu_\rho(\partial{\cal C}/\partial p_{\rho})}{\xi^\nu_\sigma(\partial{\cal C}/\partial p_{\sigma})}
=\frac{(f+gp^2)\,{\cal C}'p^\mu}{(f+gp^2)\,{\cal C}'p^\nu}=\frac{p^\mu}{p^\nu},
\ee
where ${\cal C}'=\partial{\cal C}/\partial p^2$, are functions only of the $p_\mu$ and hence both the velocity and the angles are constant due
to the equations of motion (\ref{EqMotionp}).
Hence, the spatial trajectory is a straight line. Once these conditions are satisfied, one can proceed as in previous section, even if now
$\xi=\xi(x,p)$.

It must be noticed that it is possible that more general models satisfy the conditions, at least in special cases.
For example, in the quantum clock model \cite{MignemiUras} the radial trajectories of free particles are still straight lines and the velocity is
constant along them, so our theorem is valid for such trajectories, even if the constancy of the velocity is not evident at first sight. However,
this is not true for non-radial trajectories, and for such trajectories the integration of the field equations and the action of finite translations
result too difficult to investigate, so that no definitive answer can be obtained on the identification of the physical velocity with the group 
velocity in that case.

\section{Conclusions}

We have shown that the fact that, at the level of free particles,
the group velocity coincides with the physical velocity is true in general
in the context of theories with deformed relativistic symmetries,
i.e.~of theories with (relativistic) curved momentum space. In order
to prove our result, we have implemented the effects of relative locality
for translations in a slightly different way with respect to what is done
usually in the literature. Our result holds in general
and at all orders in the deformation/curvature parameter. This clarifies
ultimately the apparent discrepancy between coordinate velocity and
group velocity characterizing these kind of theories.
When the deformation also depends on the position, our proof only holds
if the Lorentz invariance is preserved, as in generalized Snyder models,
while in more general cases (like curved spacetime) a general answer is not
known.

\section*{Acknowledgment}
This work was started during the stay of GR as Post Doc at INFN, Sezione di Cagliari, Cittadella Universitaria, 09042 Monserrato, Italy.

\end{document}